\documentclass{article}

\usepackage{arxiv}

\usepackage[utf8]{inputenc} 
\usepackage[T1]{fontenc}    
\usepackage{hyperref}       
\usepackage{url}            
\usepackage{booktabs}       
\usepackage{amsfonts}       
\usepackage{nicefrac}       
\usepackage{microtype}      
\usepackage{graphicx}
\usepackage{amsmath}
\usepackage{amssymb}
\usepackage[final]{pdfpages}
\title{ Identification and correction of Sagnac frequency variations: an implementation for the GINGERINO data analysis }
\author{Angela D. V. Di Virgilio and  Umberto Giacomelli\\
             INFN Section of Pisa, Italy
        \And
 Nicol\`o Beverini, Giorgio Carelli, Donatella Ciampini, Francesco Fuso and Enrico Maccioni \\      
 Physics dept., University of Pisa, Italy
       \And
 Antonello Ortolan\\
             INFN National Laboratory of Legnaro, LNL, Italy
 }
\begin{document}

\maketitle

\begin{abstract}
Ring laser gyroscopes are top sensitivity inertial sensors used in the measurement of angular rotation. It is  well known  that the response of such remarkable instruments can in principle access the very low frequency band, but the occurrence of nonlinear effects in the laser dynamics imposes severe limitations in terms of sensitivity and stability. We report here general relationships aimed at evaluating corrections able to effectively account for nonlinear laser dynamics. The so-derived corrections are applied to analyse thirty days of continuous operation of the large area ring laser gyroscope GINGERINO leading to duly reconstruct the Sagnac frequency $\omega_s$.
  The analysis shows that the evaluated corrections affect the measurement of the Earth rotation rate $\Omega_{\oplus}$ at the level of 1 part in  $1.5\times10^{3}$.
The null shift term $\omega_{ns}$ plays a non negligible role. 
It turns out proportional to the optical losses $\mu$ of the ring cavity, which are changing in time at the level of $10\%$ within the considered period of thirty days. The Allan deviation of estimated $\Omega_{\oplus}$ shows a remarkable long term stability, leading to a  sensitivity  better than $10^{-10}$rad/s with more than $10$s of integration time, and approaching $(8.5\pm 0.5)\times 10^{-12}$rad/s with $4.5\times10^{5}$~s of integration time.
\end{abstract}

\section{Introduction}
Ring laser  gyroscopes (RLGs) are inertial sensors based on the Sagnac effect \cite{RSIUlli,CR,DIVIRGILIOCR}. They are largely used for inertial navigation, and applications in geodesy, geophysics and even for General Relativity, where tests are foreseen \cite{brz}. Since 2011 we are studying the feasibility of the test of Lense--Thirring dragging of the rotating Earth at the level of $1\%$ with an array of large frame RLGs \cite{prd2011,angelo2017,angela2017}.
 For that purpose it is necessary to push the relative accuracy of the Earth rotation rate $\Omega_{\oplus}$ measurement in the range from $1$ part in $10^9$ up to $1$ part in $10^{12}$.
 
 \noindent RLG consists of a laser with a cavity comprising of three or four mirrors, rigidly attached to a frame; large  frame RLGs are utilised to measure the Earth rotation rate, being attached to the Earth crust.
  Because of the Sagnac effect, the two counter-propagating cavity beams have slightly different frequency, and the beat note of the two  beams is proportional to the angular rotation rate of the ring cavity. Large frame RLGs are the most sensitive instruments for inertial angular rotation measurements. The Sagnac frequency of a RLG is in fact proportional to the component of the angular velocity $\overrightarrow{\Omega}$ felt by the instrument along the normal to the cavity plane:
\begin{eqnarray}
f_s =S \Omega \cos{\theta}\\
 S = 4\frac{A}{\lambda L} \ , \nonumber
\label{uno}
\end{eqnarray}
where $A$ is the area of the ring cavity, 
$L$ is its perimeter, $\lambda$ the wavelength of the light,
and $\theta$ is the angle between the area vector of the ring and 
$\overrightarrow{\Omega}$. For RLGs lying horizontally (area vector vertical) 
 $\theta$ is the co-latitude angle, while for RLGs aligned at the maximum Sagnac signal $\theta = 0$.
 Eq.~1 defines the scale factor $S$, which is a function of
  the geometry and of $\lambda$, quantities than can be measured with a very high accuracy.
  
\noindent Further to sensitivity, other key points of such instruments rely on their broad bandwidth,  which can span from kHz down to DC,
  and their very large dynamical range. In fact the same device can record microseismic events and high magnitude
  nearby earthquakes \cite{simonelli}, being the signal based on the
  measurement of the beat note between the two counter-propagating beams.
   It has been proven that large size RLGs, equipped 
   with state of the art mirrors, can reach the relative precision of $3$ parts in $10^9$ with one day of integration time, in the measurement of $\Omega_{\oplus}$ \cite{RSIUlli}.  If shot noise limited, sensitivity level scales with the square of the size of the ring cavity \cite{RSIUlli}. However, other limitations can affect the measurement.
  
    The laser dynamics is non linear and plays a role in determining the RLG signal. We have recently developed a model to reconstruct the Sagnac frequency $\omega_s$ starting from the measured beat note $\omega_m$ and using the mono-beam signals.\footnote{For mono-beam signals we intend the signals observed by the two photodiodes that detect the laser intensities in the counter-propagating directions, see Fig.~\ref{GeneralScheme}.}\\
   The Sagnac angular frequency can be expressed as the linear sum of several terms. As we have discussed in a recent paper \cite{Angela2019}, reconstruction of the Sagnac angular frequency is mostly affected by the correction $\omega_{s0}$ associated with the so called back-scattering. In this paper we provide the necessary information to evaluate the other correction terms. Thirty days of continuous operation of GINGERINO are analysed taking into account the set of first order expansion terms, and results discussed. \\
   Results demonstrate that the so called null shift term, the first order correction $\omega_{ns1}$, plays a non negligible role in affecting both accuracy and sensitivity of the apparatus. The finding is in contrast with conventional treatments of RLG data, where null shift effects are typically neglected or considered simply proportional to the difference of the two mono-beam signals. On the contrary we found that including such correction leads to improve the measurement accuracy. Furthermore, since null shift effects can be related to the optical losses of the system, which inherently depend on operation time, a time dependent effect is found.

\section{The analysis scheme to take into account laser dynamics}
\begin{figure}[htbp]
\centering
\includegraphics[width=\linewidth]{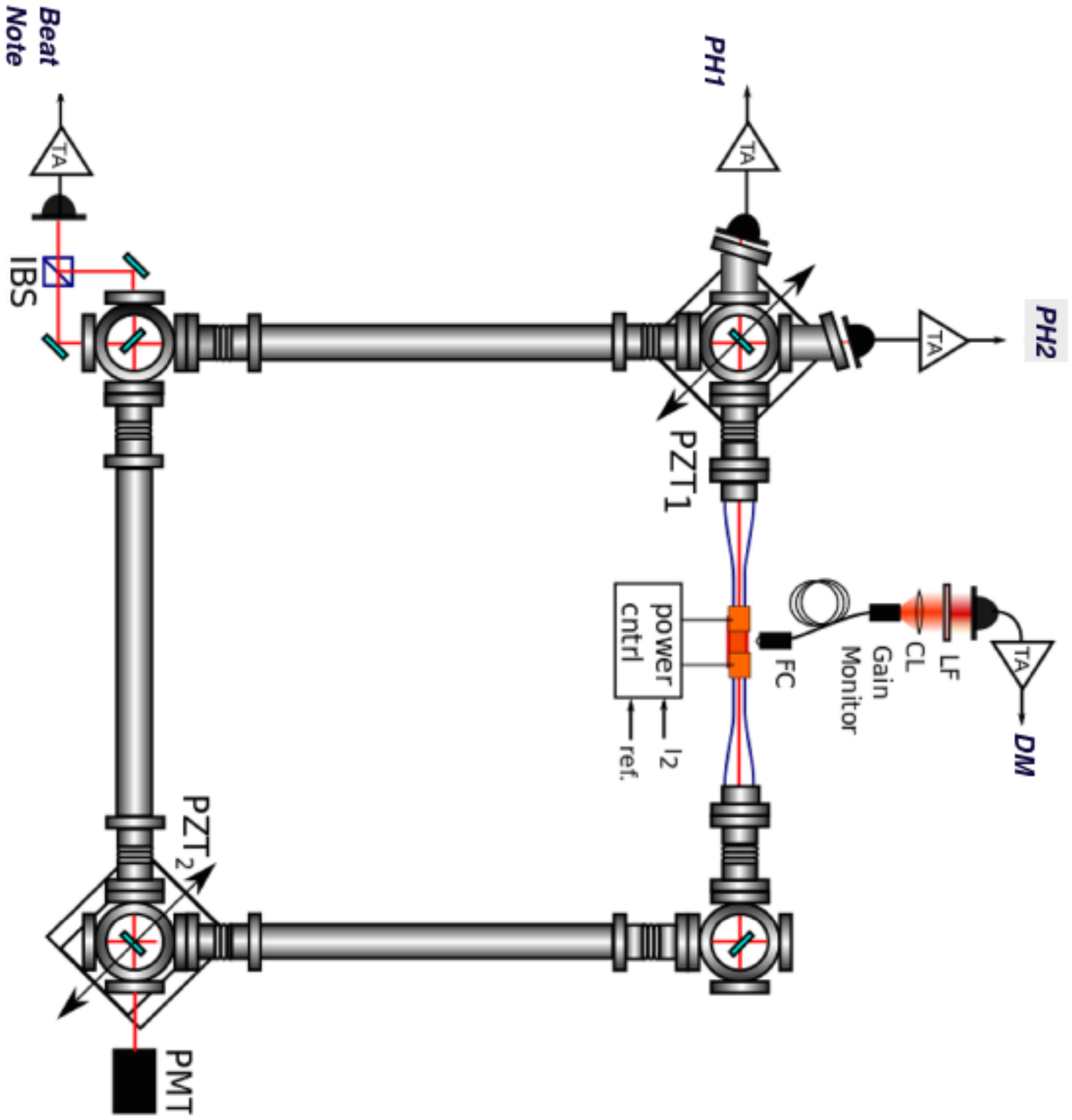}
\caption{Typical scheme of RLG with a square ring cavity. Photodiodes used to measure the beat note, necessary to evaluate the $\omega_m$, the two mono-beams ($PH1$ and $PH2$) and the gain monitor ($DM$) are shown. PMT indicates an additional photodetector not used for the purposes of the present paper. Note that the signal produced by the photodiode $PH1$ is used also in a feedback loop to control the power of the laser. }
\label{GeneralScheme}
\end{figure}

Our approach is based on the model of laser dynamics we recently developed in details in \cite{Angela2019} based on the Aranowitz model of RLG \cite{aronowitz,beghi2}, where laser dynamics is described in terms of several dimensionless parameters (Lamb parameters).
The polarization of the laser plasma is described up to the third order expansion in powers of the field \cite{aronowitz,beghi2}. 
In the case of large frame RLGs, to avoid mode competition, a 50:50 mixture of two isotopes, $^{20}Ne$ and $^{22}Ne$, is utilised, and the laser operation is set close to threshold to guarantee single-mode operation. This particular choice allows for some simplifications: the Lamb parameters of cross-saturation  ($\theta_{12}$ and $\theta_{21}$) can be  neglected, and we can assume the self-saturation terms equal to each other (i.e., $\beta_1 = \beta_2 = \beta$).\\
The general goal of the analysis is to evaluate the Sagnac frequency correcting systematics due to the laser nonlinear dynamics by using the available data: the measured beat note and the mono-beam signals,
\cite{Angela2019,90day}. Figure \ref{GeneralScheme} shows a scheme  of the experimental apparatus, where photodiodes collecting the relevant signals are indicated. 

\noindent
The model provides two analytical expressions for the Sagnac angular frequency $\omega_s = 2 \pi f_s$: a non-linear relationship (Eq. 8 of ref. \cite{Angela2019}), which connects all the laser parameters with each other, and an approximate decomposition of $\omega_{s}$ as a linear sum of several terms, which collects the first and second order expansion of the correcting terms (a total of six terms):
\begin{equation}
\omega_s = \omega_{s0} +\omega_{ns1} +\omega_{ns2}+\omega_{K1} +\omega_{K2}+\omega_{nsK}.
\label{eq:uno}
\end{equation}
As demonstrated in \cite{Angela2019} $\omega_{s0}$, accounting for back-scatter effects, is the dominant term in the expansion.
The term $\omega_{s0}$ does not depend on Lamb parameters and can be expressed starting from the knowledge of experimental parameters according to:
 \begin{eqnarray}
\omega_{s0} = \frac{1}{2} \sqrt{\frac{ 2 \omega_m^2 I_{S1}   I_{S2} \cos (2 \epsilon )}{  I_{1}  I_{2}}+\omega _m^2}+\frac{\omega _m}{2} + \omega_{s \xi}\\
\label{formula}
\omega_{s \xi} = \xi\times\frac{  I_{S1}   I_{S2} \omega_m ^2 \cos (2 \epsilon )}{2  I_{1}  I_{2} \sqrt{\frac{ 2  I_{S1}   I_{S2} \omega _m^2 \cos (2 \epsilon )}{ I_{1}  I_{2}}+\omega _m^2}}
\label{xi}
\end{eqnarray}
where $\omega_m$ is the measured beat note (expressed as angular frequency), $I_{1,2}$, $I_{S1,S2}$, and $\epsilon$ are  the DC level of the mono-beam signals, their amplitude demodulated at the beat note frequency, and the dephasing between the two signals, respectively.\\
\noindent

In order to better define $\omega_{s0}$  and take into account experimental issues, such as dark currents in the photodiode signals, the term $\omega_{s\xi}$, see Eq. \ref{xi},  has been introduced, where $\xi$ is  a proportionality constant which could be evaluated by statistical means.\\
 The other terms in Eq. \ref{eq:uno} give smaller contributions. They can be evaluated as detailed in the Appendix, starting from the  knowledge of the Lamb parameters \cite{aronowitz}, a careful estimation of all operating parameters of the system, and the evaluation of the total optical losses $\mu$. The latter can be measured with a $1\%$ accuracy via the ring down method \cite{beghi2}, i.e., by suddenly switching off the laser discharge and measuring the time behaviour of the intracavity signals through photodiodes $PH1$ and $PH2$ in Fig.~\ref{GeneralScheme}. The method cannot be applied continuously during the RLG operation, since it requires to switch off the laser. Moreover, optical losses can change during operation, therefore their evaluation should be repeated frequently.
 \\
 Because of that, and also to avoid uncertainties and fluctuations related to estimation of operating parameters such as, laser beam size, pressure and polarization conditions of the discharge (see Appendix for their use in the relevant expressions), in the present paper we have adopted a statistical approach for evaluating $\mu$. As shown in the next subsection, this parameter enters as a linear proportionality factor into $\omega_{ns1}$. Therefore, we can express $\omega_{ns1} = \mu \times \tilde{\omega}_{ns1}$, where $\tilde{\omega}_{ns1}$ is evaluated starting from known data and $\mu$ determined through a linear regression procedure. Although the  other terms in the expansion of Eq. \ref{eq:uno} are less relevant, a similar approach is also applied to terms containing the subscript $K$, expressed through an explicit multiplication by a factor $MK$ which is evaluated via the statistical approach. 
 
\begin{figure}
    \centering
     \includegraphics[scale=0.3, angle=90]{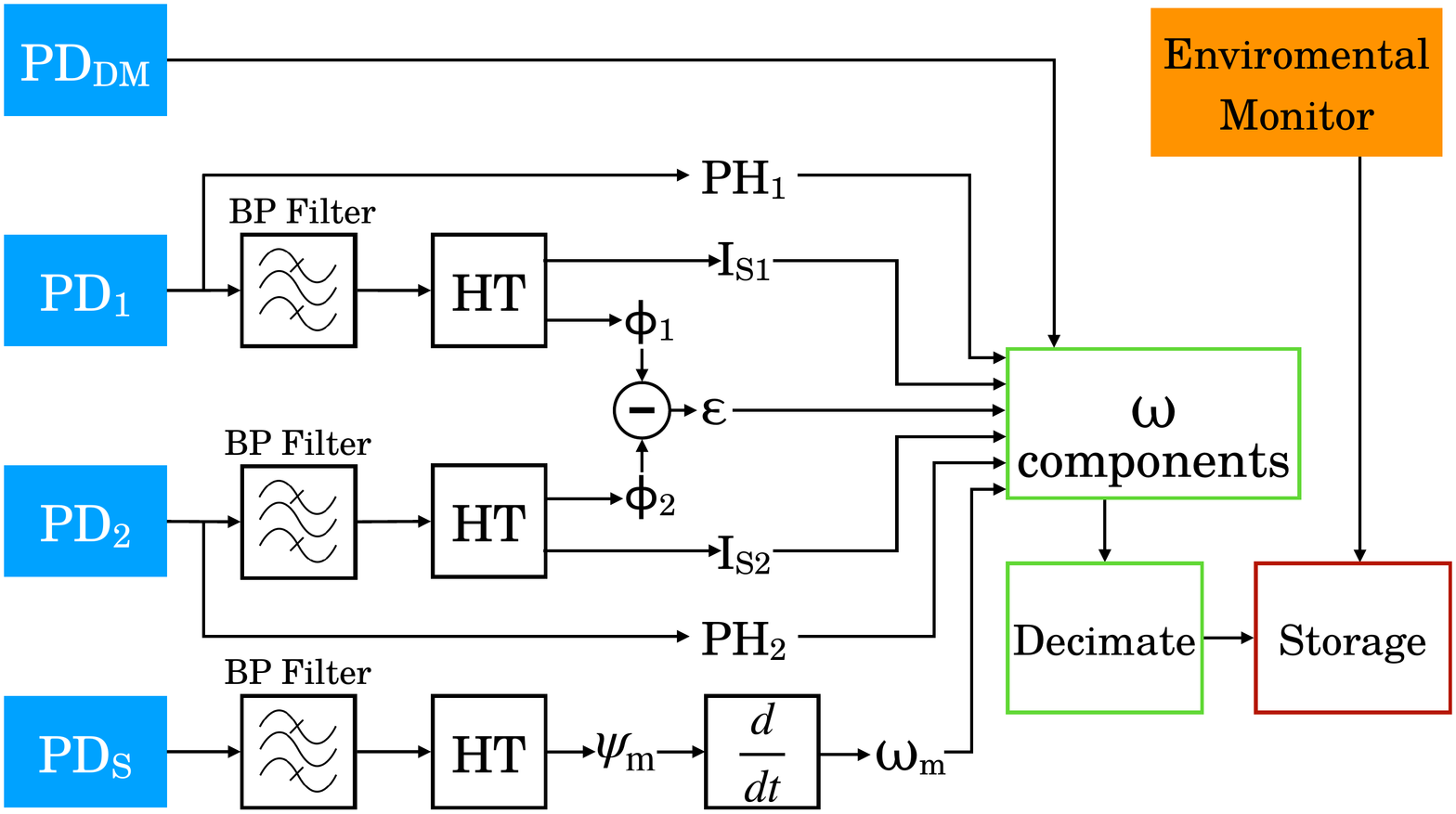}
     \caption{First analysis level and data storage. Symbols $PD_{...}$ are used in this figure to denote different photodetectors, BP stands for band-pass, HT for Hilbert transform. }
     \label{fig:diagram}
 \end{figure}
\begin{figure}
     \centering
     \includegraphics[scale=0.14]{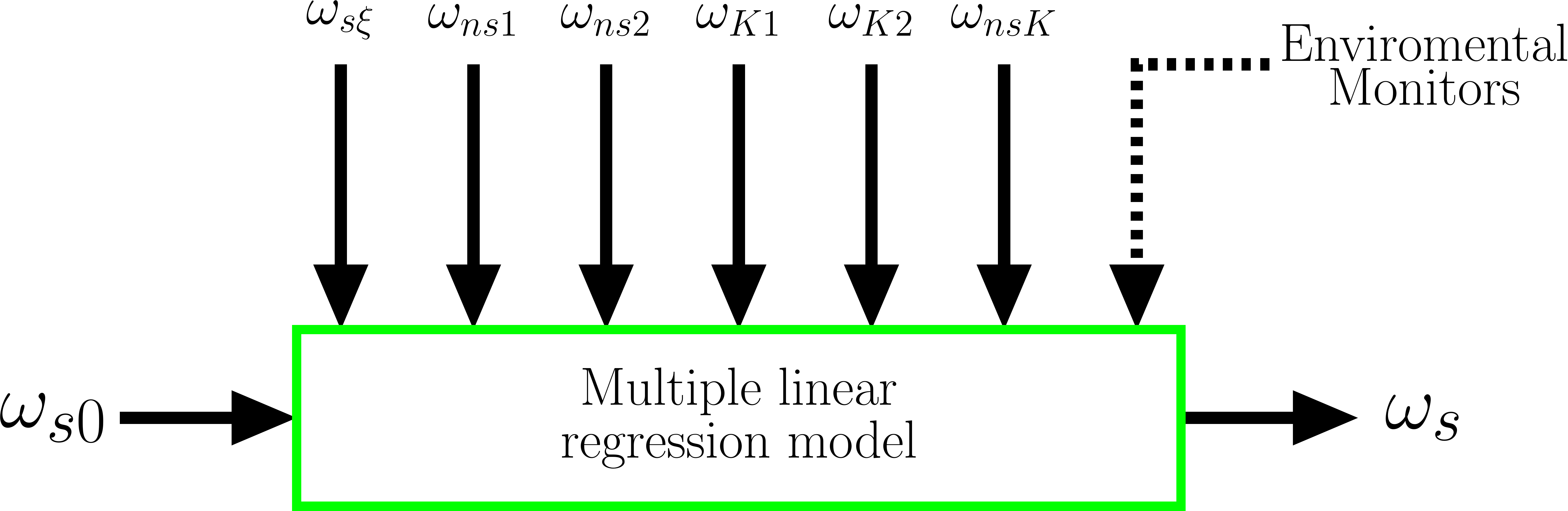}
     \caption{Linear regression model analysis scheme:  parameters $\xi$, $\mu$, and $MK$ related to the reduction of systematics due to laser dynamics are evaluated by statistical means in order to determine $\omega_{s}$. Other parameters coming from environmental monitors could also be included in the statistical approach.}
     \label{fig:diagram2}
 \end{figure}

\noindent
In summary, the analysis procedure is as follows: in the first level, different terms $\tilde\omega_{ns1,2}$ are evaluated and stored on disk; in the second level,  updated $\xi$,  $\mu$ and $MK$ coefficients are found with linear regression models, minimising the sum of Eq.~\ref{eq:uno}. It is important to remark that $\mu$ can be stable with time, or varying with time. 
In this second case it is necessary to add to the model the information necessary to follow the time behaviour. Subsection \ref{sec:step2} presents the method used for GINGERINO.
We remind that the available signals from the RLG are sampled at high frequency (typically 5 kHz), and
the whole calculation is carried out at sufficiently high frequency, in our case 2.5 kHz, while data are stored after decimation, usually corresponding to a sampling frequency of 2, 1, or 0.1 Hz. The diagram in Fig. \ref{fig:diagram} shows the first level involved in the method, while Fig. \ref{fig:diagram2} presents a diagram of the second level of analysis.
 
 \noindent
 In the following, 30 days of data produced by GINGERINO will be analysed. In the analysis, corrections are included only up to the first order: $\omega_{ns1}=\omega_{ns}$, $\omega_{K1}=\omega_{K}$. Second order terms are in fact smaller and do not contribute to improve the results compared to the present state of accuracy. The Appendix contains however all information to evaluate the different terms in the general case.
 

\subsection{Some details on the evaluation of first order expansion terms}

 As already mentioned, we start from the DC components, $I_1$ and $I_2$, of the mono-beam signals, their AC components at the beat frequency $\omega_m /2 \pi$, $I_{S1}$ and $I_{S2}$, and the relative phase $\epsilon$ between the two AC components.\\
Dissipative processes (like  diffusion or absorption from the mirrors) can produce non reciprocal losses between the two counter-propagating beams, so that two distinct loss parameters $\mu_{1} \neq \mu_{2}$ must be used.
 Without loss of generality, it is possible to take $\mu_1=\mu$ and  $\mu_2=\mu+\delta\mu$, where $\mu$ represents the reciprocal losses term and $\delta\mu$ the non reciprocal ones. 
The plasma dispersion function depends on several parameters: gas pressure, line width of the excited isotopes, resonance frequency of the two counter propagating beams, area of the beam profile at the discharge, and temperature of the plasma.  
Assuming that $\beta_1 = \beta_2 = \beta$ (a rough evaluation reported in \cite{beghi1} gives $\beta_1 - \beta_2 \simeq 10^{-14}$), combining the information of the mono-beam DC signals $I_1$ and $I_2$ with the plasma dispersion function, the gains of the two beams Gain1 and Gain2 (see Appendix) are evaluated. Since $\beta_1 = \beta_2$
the non reciprocal loss term
$\delta\mu$ can be hence determined. 
Using other relationships reported in the literature \cite{cuccato} it is possible to write $\omega_{ns1,2}$, $\omega_{K1,2}$ and $\omega_{nsK}$ as a function of the known quantities $\omega_m$, $I_1$, $I_2$, $I_{S1}$, 
$I_{S2}$, $\epsilon$, and $\mu$.
As shown in the Appendix, losses appear as a multiplicative factor in $\omega_{ns1,2}$.\\

\noindent 
Further to be dependent on losses, terms correcting the null shift depend also on $\epsilon$, as well as on geometry and size of the cavity, in a rather complicated way. Therefore, minimization of $\omega_{ns1,2}$ is not straightforward.
To give an example using the parameters of GINGERINO, the minimum is found for $\delta\mu = 0$; assuming typical values in physical units $I_1=I_2=1$ V and $I_{S1}=I_{S2}=0.01$ V (see Appendix for the calculation), the null shift correction is of the order of 0.4 $\mu$Hz, with a modulation induced by variations of $\epsilon$ of the order of $5\times10^{-6}$ $\mu$Hz.
In general $\delta\mu \neq 0 $, as can be inferred from the difference in mono-beam signals. While keeping all other parameters fixed as in the example above, a difference in the demodulated photodiode signals $I_{S1}$ and $I_{S2}$ of 1 mV leads to remarkable values of the null shift correction. In such conditions, the null shift contributes to around 500 $\mu$Hz, with modulations induced by $\epsilon$ of the order of 0.013 $\mu$Hz. \\
\noindent 
The above examples suggest that, in general, variations with $\epsilon$ are rather small and can be neglected. Therefore, in the ideal case of equal losses the effect of the null shift is a systematic term affecting the accuracy of the measurement at the level of $1.5\times 10^{-9}$, assuming variations of the losses with time below $10\%$, while when $\delta\mu$ is not negligible the effect is orders of magnitude higher. In the considered case the requirement on the stability of $\mu$ with time is of the order of 1 part in $10^4$ in order to achieve a precision of 1 part in $10^9$. As far as accuracy is concerned, the systematic effect is of the order of 1 part in $10^6$, one thousand time  the goal of GINGER (1 part in $10^9$ at least).\\
While being less relevant than those related to null shift effects, also the terms with the $K$ suffix deserve discussion. 
Eq.~\ref{eq:uno} shows that such terms are linearly summed to $\omega_s$. As a consequence, their contribution to low frequency signals, i.e., those stimulating the major interest in typical applications of RLGs, can be reduced by low pass filtering. 
Despite of the possibility to remove related contributions through low pass filtering, the evaluation of $\omega_K$ requires some further assumptions. $\omega_K$ 
oscillates at the Sagnac frequency, containing  $\sin(\omega_s t)$ and $\cos(\omega_s t)$ terms.  
However, due to nonlinear dynamics, the long time average of such oscillating terms can deviate from zero. Therefore, we can express $\omega_K$ as the product of a small parameter $MK$ with a given function of  known parameters. As already mentioned, the value of $MK$ and its very slow variations can be found via linear regression. \\
\noindent
A Mathematica notebook is reported in the Appendix to show details of the symbolic calculations leading to $\omega_{ns}$ and $\omega_K$ functions, to demonstrate how losses are a scale factor for the $\omega_{ns}$ terms, and to provide numerical examples of the involved values calculated for the GINGERINO setup. We note that, according to our model, $\omega_K$ can be written as the ratio of two polynomials where the parameter $\mu$ enters in one term, preventing direct evaluation of such a parameter through a simple fit procedure. However, we have checked that, with typical $\mu$ values of $100-1$ ppm,  variations in calculated results are not significant. \\
In the following, we focus onto calculations for a general RLG dynamics, and their implementation for the specific case of GINGERINO.\\
\subsection{The selected playground and the first step of the analysis}
Thirty days of continuous operation between June 16 and July 15 2018 have been selected for this test. After June 21, heavy operations took place in the underground laboratories where GINGERINO is placed for the building of another experiment. Disturbances induced in GINGERINO operation, consisting of mostly mechanical effects, are well visible in the second half of the data stream. We included also these data to show that, even in non ideal conditions, very high sensitivity and stability can be obtained.
The data are acquired with our DAQ system at 5 kHz sampling rate\cite{90day}. The whole set of data has been analysed on a hourly base, and the terms $\omega_m$, $\omega_{ns}$ (with $\mu=1$), $\omega_\xi$ (with $\xi=1$), and $\omega_K$ (with $MK=1$ and $\mu=10^{-4}$) evaluated and stored for offline analysis. In the calculation, average losses $\mu=10^{-4}$ are used to evaluate $\omega_K$, since it has been checked that the dependence on $\mu$ is negligible at the present sensitivity limit. The analysis, based on the Hilbert transform of interferograms and mono-beam signals, leads to the beat note frequency  $\omega_m /2 \pi$, the amplitudes of the mono beams at the beat frequency $I_{S1}, I_{S2}$, and the relative phase $\epsilon$.
In order to avoid spurious oscillations due to the boundaries between contiguous hours in the hourly-based analysis, for each hour 6 minutes of data are added at the beginning and at the end; these extra samples are removed after the analysis (overlap-save method). 
The analysed data are stored after decimation at 20 and 0.1 Hz; the decimation is implemented via the standard Matlab function decimate, which applies 8th order Chebyshev Type I lowpass filter with cutoff frequency $0.8\times(Fs/2)$, where $Fs$ is $20$ or $0.1$ Hz. 

\subsection{The second step of the analysis}
\label{sec:step2}
 A second routine collects offline different days for longer analysis. The first operation is to identify the portions of data which are not at normal operation, typically less than $5$ \% of the data are removed. To this aim, we make use of the fringe contrast of the interferogram acquired at $20$ Hz rate. The choice of such a sampling rate minimizes the discarded sample. This operation associates the flag 1
 to the good samples and the flag 0 to the bad ones; from this routine we get the mask to select data decimated at lower frequency.\\
 In the present analysis, decimated at half an hour rate, above $90\%$ of data are selected. 
 Both the intensity of the mono-beam signal ($I_2$, collected by the photodiode indicated as $PH2$ in Fig.~\ref{GeneralScheme}), that is not used for the feedback control of the laser power, and of the Discharge Monitor ($DM$, i.e. a photodiode looking at the laser discharge) change
with time.  This is a clear indication that $\mu$ changes with time; in particular, $I_2$ indicates the presence of non reciprocal losses $\delta\mu$,
while $DM$ is proportional to the total excited atoms, and therefore 
to the total losses. $DM$ and $I_2$ signals allow us to model the change of losses as a function of time through
 two form factors: 
 \begin{itemize}
     \item gain monitor form factor: $FF_{DM} = \frac{DM-<DM>}{<DM>}$ 
     \item mono-beam 2 form factor: $FF_{PH2}=\frac{I_2-<I_2>}{<I_2>}$.
 \end{itemize}
\subsection{Reconstruction of the Sagnac frequency by a linear regression model}
\begin{figure}
    \centering
    \includegraphics[scale=0.13]{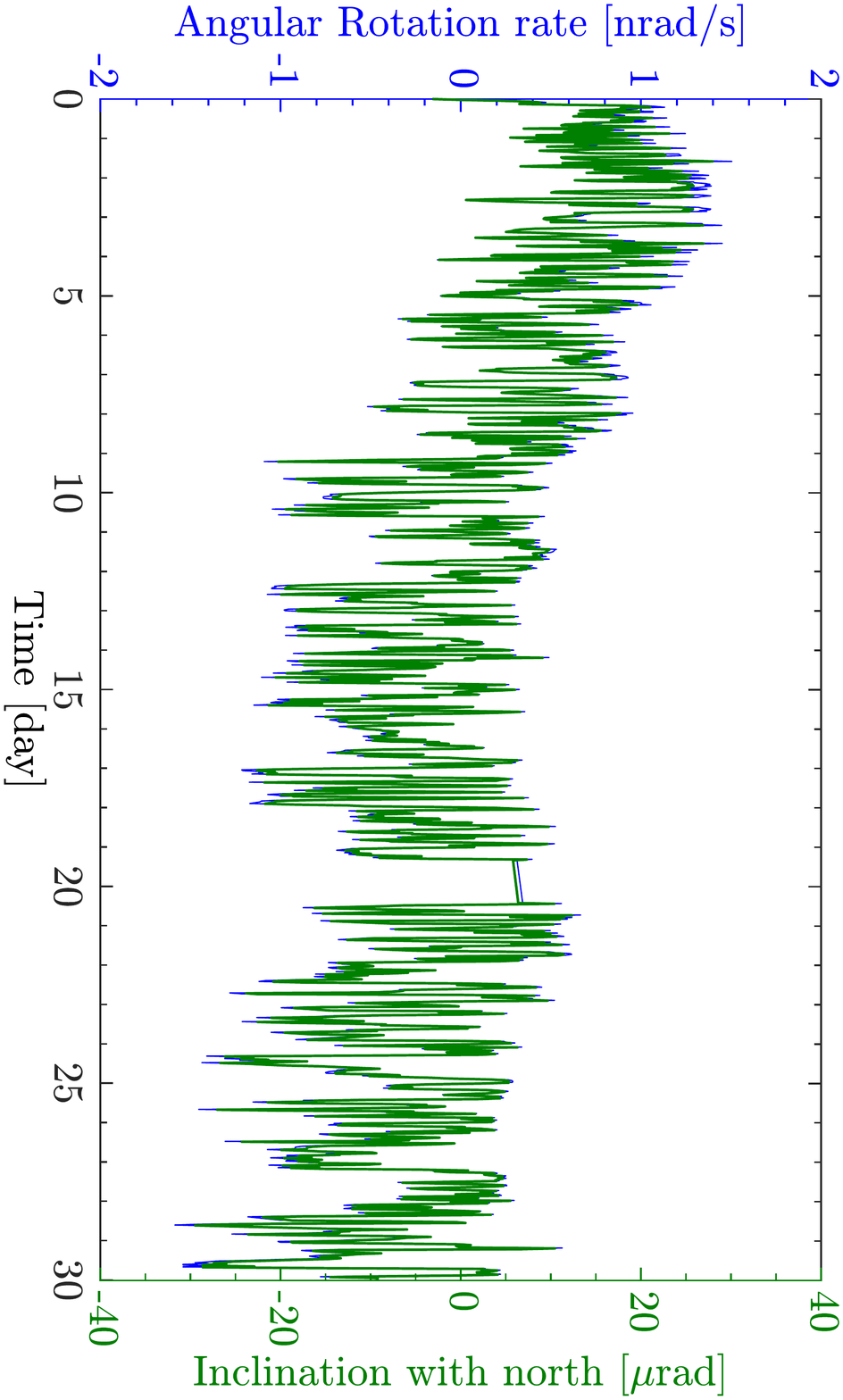}
     \includegraphics[scale=0.13]{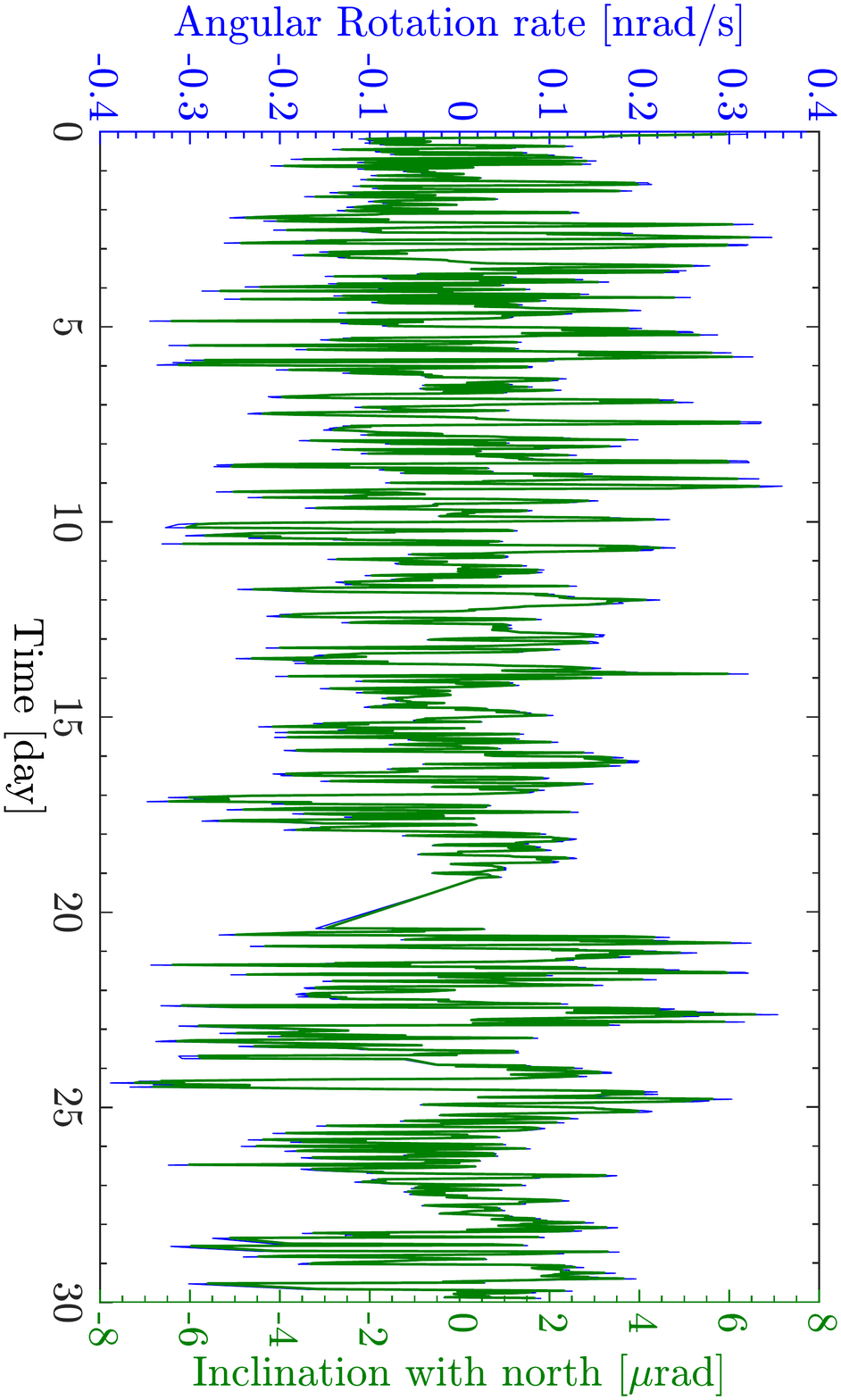}
    \caption{Top: Time variations expressed as $\Omega$ and $\theta$, see Eq.~\ref{uno}, of the analysed data [utilising  $\omega_{s0}$ with the mean value subtracted; the mean value is $2 \pi \times (280.208\pm 0.001$ Hz), compatible with a RLG with area vector vertical within a few mrad error]. 
    Bottom: as above, but using $\omega_s$ evaluated with the linear regression model.
    Note: around day 20 some data were lost because of a failure in the data acquisition system.}
    \label{fig:histro1}
\end{figure}

The vectors $\omega_{ns}$, $\omega_{ns}\times FF_{DM}$, $\omega_{ns}\times FF_{PH2}$, $\omega_{\xi}$, and $\omega_{K}$ are collected, and the linear regression method is utilised to determine the unknown parameters $\mu$, $MK$ and $\xi$, and evaluate $\omega_s$ accordingly.\\
Let us remind that $\omega_{s0}$ accounts for back-scatter noise,
and an additional term $\omega_\xi$ has been implemented to account for inaccuracies on the measured quantities $I_{S1,2}$, $I_{1,2}$ and $\epsilon$\cite{Angela2019}.
In summary, in the procedure $\mu$ and $MK$ are physical quantities which could be evaluated independently, while $\xi$ compensates noises in the measured quantities.\\
Fig.~\ref{fig:histro1} shows data before (top panel) and after (bottom panel) the application of this procedure. The dispersion across the mean value is clearly suppressed using $\omega_s$, meaning that the new regressive analysis, which includes the terms $\omega_{ns}$ and $\omega_K$, duly accounts for it.
The Matlab function {\sl fitlm} has been used to perform linear regression. 
Results indicate that the p-value, which tests the null hypothesis, is approximately zero for all vectors excluding the one related to $\omega_K$, whose p-value is 0.038. On the other hand, the relative error in the estimation of the $MK$ term involved in the $\omega_K$ expression is around 50\%, whereas it amounts to $1-12$\% for all other vectors. \footnote{The coefficient of determination of the linear regression procedure and the standard deviation of the obtained residuals are $R^2 = 0.952$ and $(770\pm3)$ $\mu$Hz, respectively.} 
 
\begin{figure}
    \centering
    \includegraphics[scale=0.24]{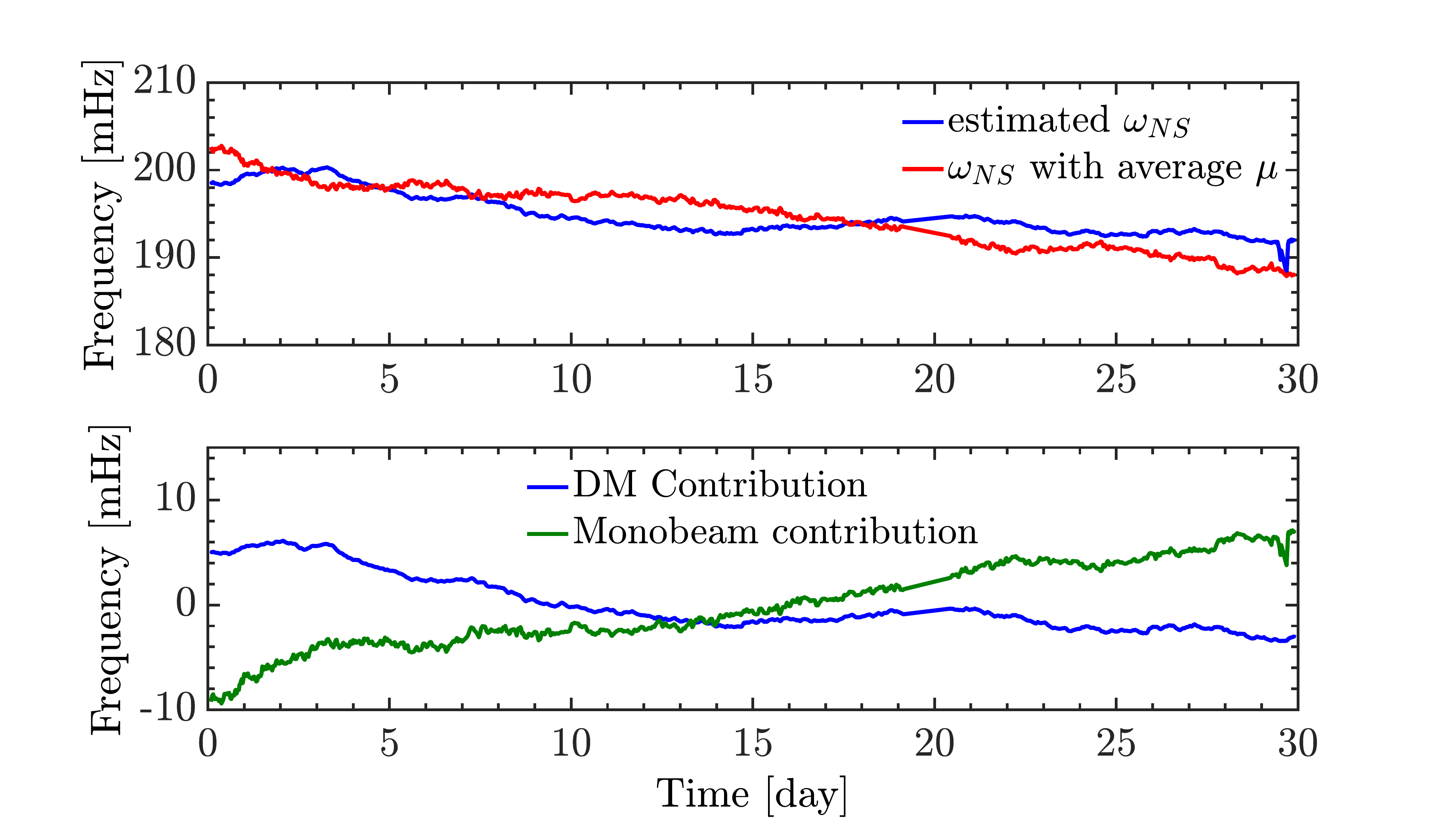}
    \caption{Top: $\omega_{ns}$ evaluated through the best-fit procedure
    and the one calculated assuming average losses.
    Bottom: the contribution using the $DM$ and $PH2$ signals to follow the time behaviour of the losses $\mu$.}    
    \label{fig:LaserSYS}
\end{figure}

\begin{figure}
    \centering
    \includegraphics[scale=0.22]{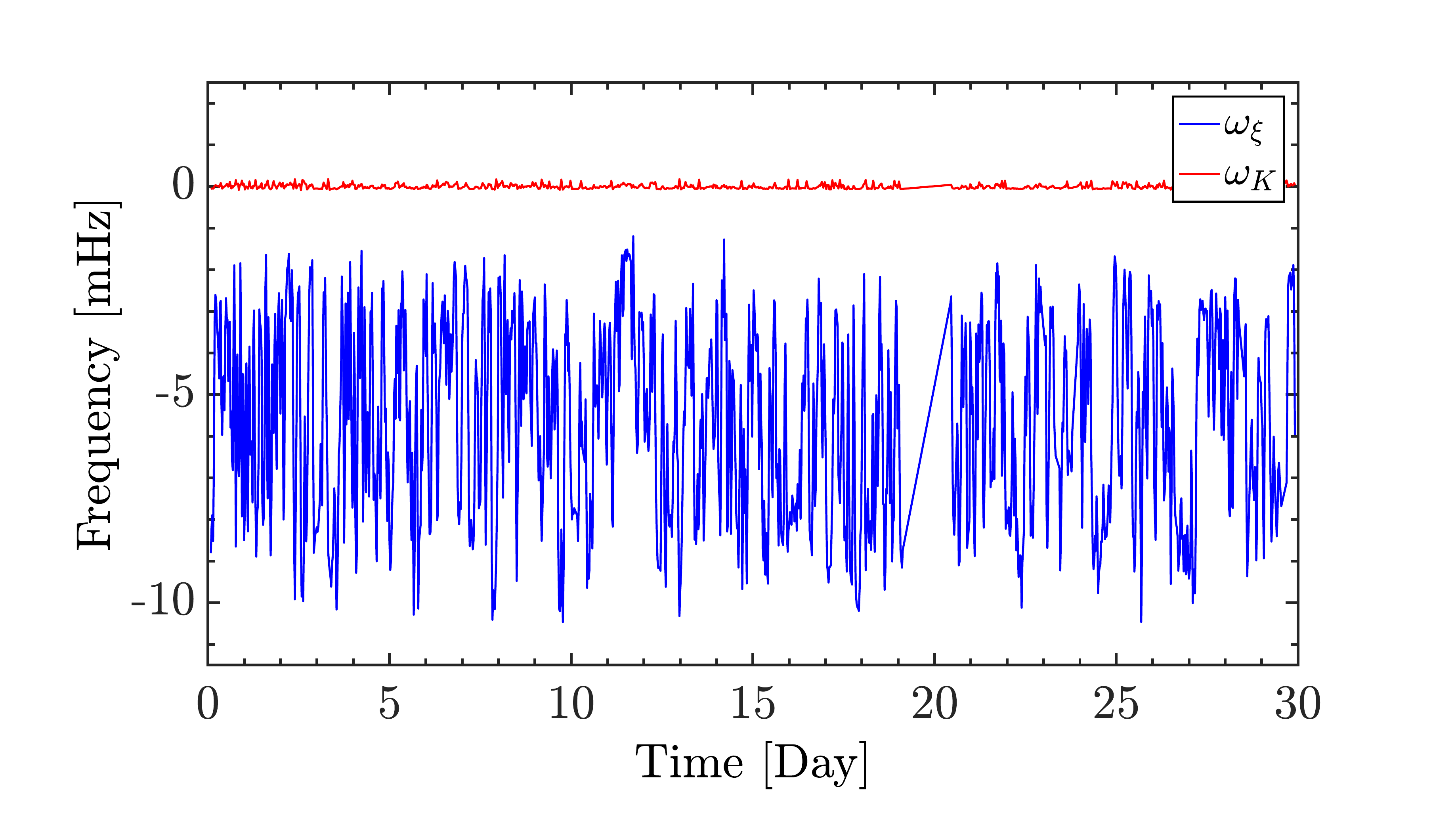}
    \caption{The first step of the analysis evaluates 
    $\omega_{s0}$, which accounts for back-scatter noise;  $\omega_\xi$ has been developed to further improve back-scatter cancellation, taking into account inaccuracies in the signals produced by photodiodes $PH1$, $PH2$, their demodulated intensity $I_{S1,2}$ 
    and dephasing $\epsilon$.}
    \label{fig:CSI}
\end{figure}

 \noindent Fig.~\ref{fig:LaserSYS} shows the contribution of the laser systematics. A dominant role is played by $\omega_{ns}$, which produces a slowly variable level ranging, in units of frequency, between $180-200$ mHz, with standard deviation $2.3$ mHz. $\omega_\xi$ gives a small correction with mean value $-5.7$ mHz and standard deviation $2.3$ mHz. These two terms affect residuals, whereas $\omega_K$ is smaller by more than a factor 10; it has a mean value compatible with zero and standard deviation $0.05$ mHz, see Fig. ~\ref{fig:CSI}.
 
\begin{figure}
    \centering
    \includegraphics[scale=0.22]{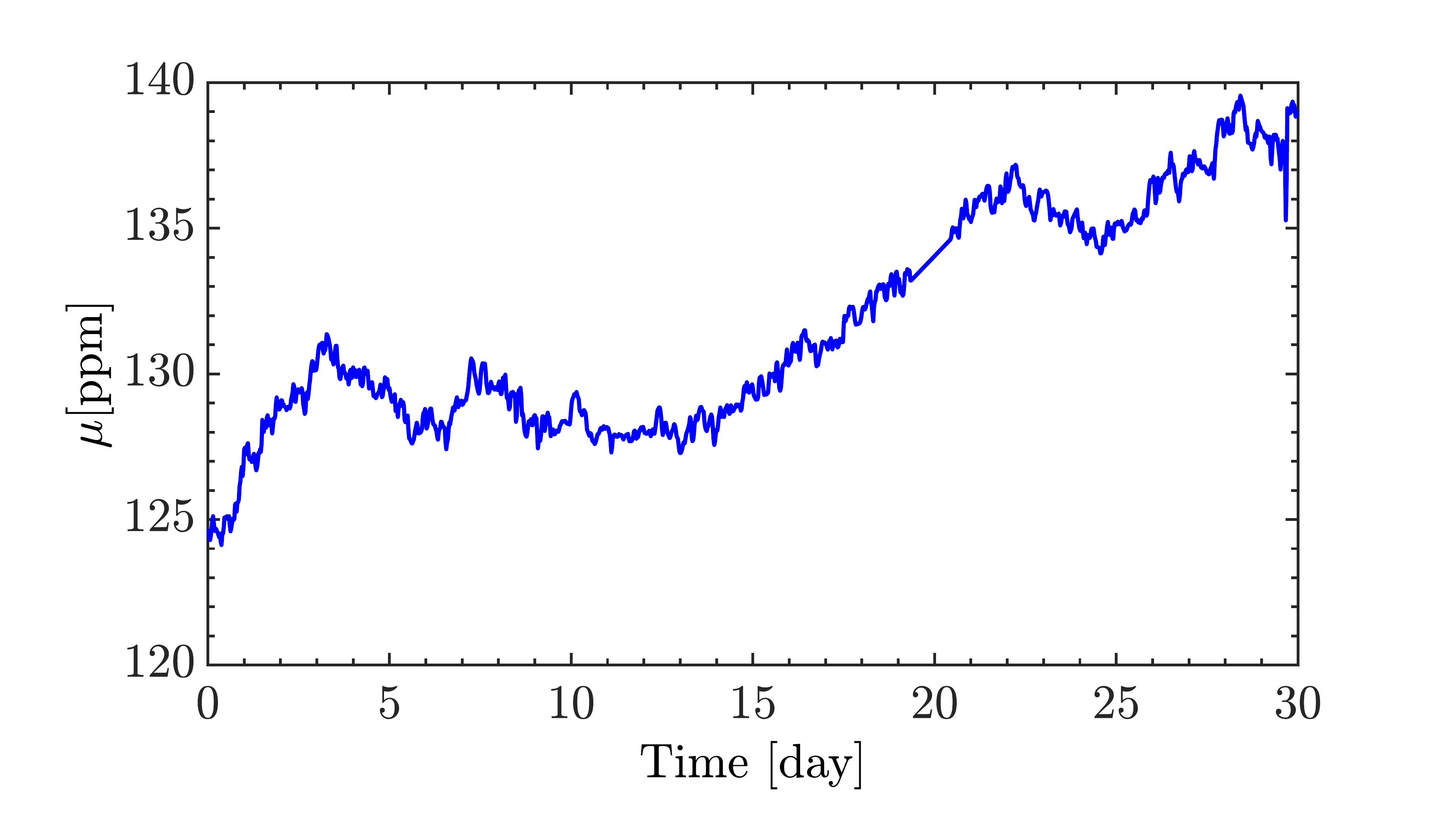}
    \caption{Time behaviour of the losses, estimated
    with the $DM$ and the uncontrolled mono-beam signal $PH2$.}
    \label{fig:mu}
\end{figure}

\begin{figure}
    \centering
    \includegraphics[scale=0.22]{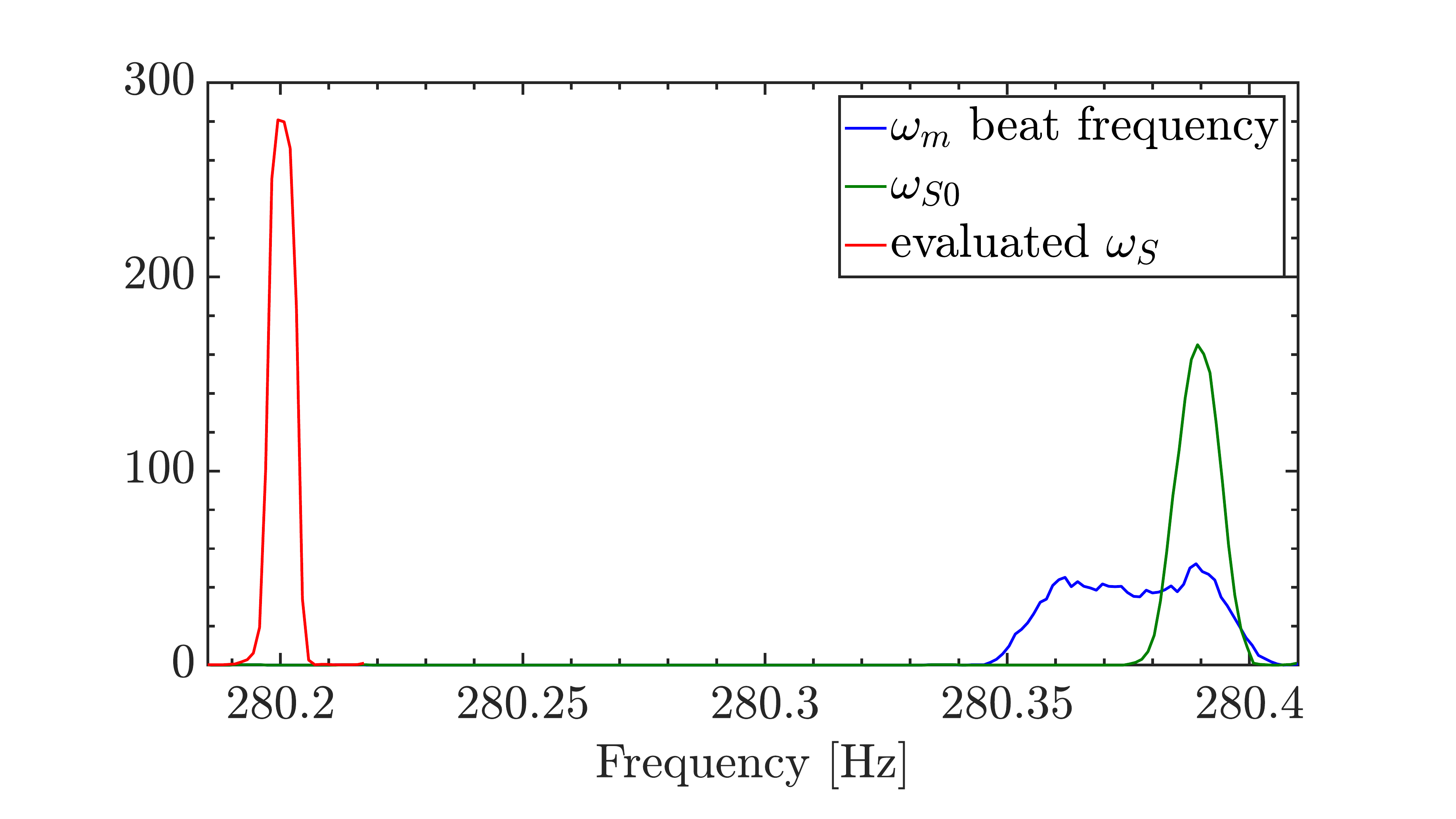}
    \caption{Distribution of the beat frequency $\omega_m/ 2 \pi$,
    of the first stage of the reconstruction $\omega_{s0} /2 \pi$ and of the final $\omega_s /2 \pi$.}
    \label{fig:histro}
\end{figure}

\begin{figure}
    \centering
    \includegraphics[scale=0.22]{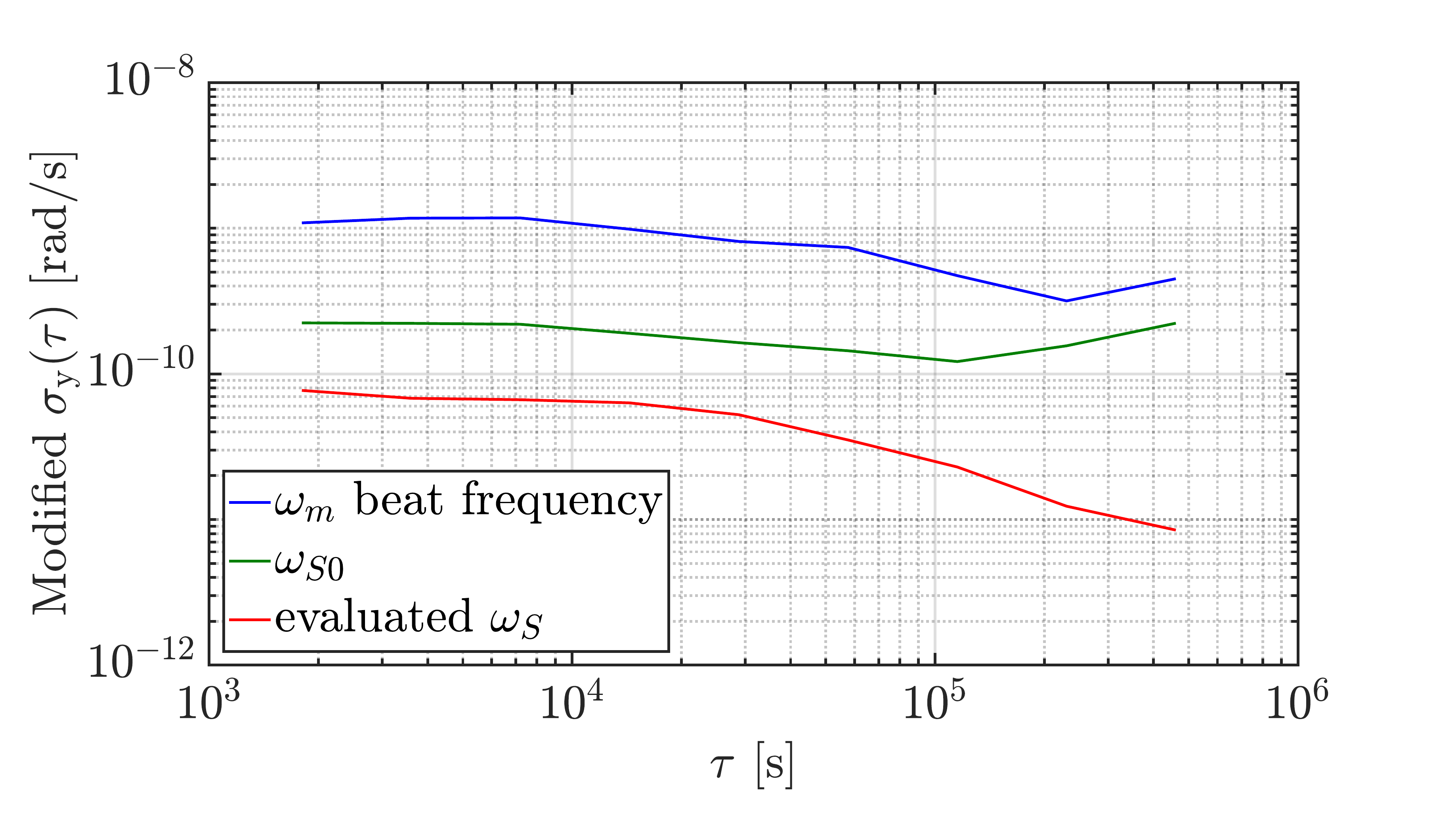}
    \caption{Modified Allan deviation of the measured angular velocity $\Omega$ from the beat note ($\omega_m$), $\omega_{s0}$ and evaluated $\omega_s$ (blue, green and red lines, respectively), relative to the mean value, expressed in angular velocity  [rad/s].}
    \label{fig:Allan}
\end{figure}

\noindent Fig.~\ref{fig:mu} shows the evaluated losses. During the $30$ day period considered, a variation of $12\%$  is observed in $\mu$, with a visible trend towards higher losses for increasing time.
Fig. ~\ref{fig:histro} compares the distributions of the beat frequency $\omega_m /2 \pi$,
of the  $\omega_{s0} /2 \pi$ reconstructed in the first stage of our approach and that of the finally determined $\omega_s / 2 \pi$. It should be noticed  that the distribution of $\omega_m$ is highly non-Gaussian while that  
of both reconstructed frequencies becomes close to Gaussian, indicating that the nonlinear terms of dynamics are correctly accounted for. Moreover, $\omega_s$ is shifted towards lower frequencies as a consequence of the null shift term $\omega_{ns}$.
\\
\noindent Fig. ~\ref{fig:Allan} reports the modified Allan deviation, expressed in angular rotation rate. We observe that, after the application of the procedure (red line), the variance is decreasing with the integration time up to more than two days. This suggests that  with our procedure we are effectively correcting the long term laser dynamic effects.
It is well known that RLGs are sensitive to Chandler Wobbler effect, 
which is typically below $100$nrad. We have checked that the level reached in the long term stability is above the limit imposed by the Chandler Wobbler effect. 
The so far obtained long term instability is about $10$ times larger than the scale factor changes induced by  temperature variations considering $5\times10^{-6}$/\textsuperscript{o}C the thermal expansion coefficient of granite, the material used for the RLG frame, and  $0.02$\textsuperscript{o}C as typical temperature variation in the 30 day period.
The obtained modified Allan deviation represents a fair improvement compared to the analysis in \cite{90day} (the improvement is around a factor 7) carried out by using available signals on a pure phenomenological basis, for instance the long time behaviour was cancelled using the $DM$ signal.\\

\section{Discussion and Conclusions}
The Sagnac frequency $\omega_s$ of the GINGERINO RLG has been evaluated taking into account the laser dynamics. The model leads to a linear sum of several terms, which can be determined with the available signals of beat frequency and mono-beams.
The so-called back-scatter noise, which has been so far considered the most severe limitation of RLG, is accounted for by terms $\omega_{s0}$ and $\omega_\xi$. It turns out that the null shift term $\omega_{ns}$ is the second dominant contribution. It depends linearly on cavity losses $\mu$;  this term definitely affects accuracy of the apparatus, since in the best case, if the losses are constant, it represents a non-negligible DC contribution. Furthermore, it is likely that losses are not constant with time, hence $\omega_{ns}$ affects stability. 
If a long term stability of 1 part in $10^9$ is required for the $\Omega_{\oplus}$  measurement, the requirement for the stability of $\mu$ with time, or equivalently for the relative accuracy in the evaluation of $\mu$, is $\frac{\omega_{ns}}{\omega_{s}}\times 10^9$. According to the model, it is straightforward to see that the null shift level is minimised when $\delta\mu\simeq 0$.
\\
\noindent Changes of $\mu$ at the level of $10\%$ are evident in GINGERINO. In order to follow those changes we have utilised the DC mono-beam signal of the uncontrolled beam, that is the signal which is not used in feedback loops controlling the laser operation (beam 2 in our case), and the discharge monitor $DM$, also called gain monitor. 
$\omega_s$ is evaluated with a linear regression model.
 \\
 \noindent Remarkably, during the measurements considered in this paper, in particular after the twentieth day of the 30 day record, 
 heavy activity was present for the construction of a new experiment inside the underground laboratory hosting the RLG. 
 Despite that, and the fact that GINGERINO is not equipped with control of the geometrical scale factor (accordingly its wavelength is not fixed), in $5.4$ days of integration time it has reached the relative stability of $(1.7\pm0.07)\times10^{-7}$
[corresponding to a resolution in the angular velocity measurement of $(8.5 \pm 0.5)\times 10^{-12}$rad/s]. \\
\noindent
The analysis poses the problem of the accuracy of angular velocity estimations;
however, it is not straightforward to understand the accuracy of the final measurement, an issue to be addressed by future
Monte Carlo studies to be carried out following the approach we have already employed in \cite{beghi1}.\\
Main goal of the present analysis was to demonstrate that null shift effects can be evaluated in analytical terms and that they are not negligible at the present sensitivity level of our instrumentation. 
In order for the linear regression method to give reliable results, the model must be complete. Our data are affected by seismic contributions. They are obviously not considered in the model, but, due to their typical frequency, their role is suppressed thanks to low pass filtering used in our procedure. Furthermore, environmental conditions, in particular temperature, can affect RLG operation. We have verified that their role is negligible for the present data set.
In order to check that the procedure does not artificially affect  shape and properties of the input signals, we have  added a spurious signal synthetically produced to the data, with amplitude of the order of fractions of nrad/s, and verified that it was not modified due to the procedure.

\noindent
The analysis has also shown that losses are the main limitation of GINGERINO. The question is now which part of the apparatus is the main responsible for losses and for their behaviour as a function of time. The hypothesis is that the main contribution comes from the  gain tube of the laser. In the standard high sensitivity RLG scheme, it acts both as gain tube and spatial filter. The laser discharge is in fact produced within a  pyrex capillary, with inner diameter of $4$ mm, which is a bit small compared to the waist of the beams. \\
In the near future, a new capillary tube with slightly larger inner diameter will be installed in GINGERINO to limit losses. Furthermore, additional data, such as current of the discharge tube and laser wavelength, will be continuously recorded in order to better characterise the instrument operation in the long term.\\
As far as the analysis is concerned,  the effect of terms $\theta_{1,2}$ will be eventually included in the model to account for possible deviations of the laser mixture behaviour from the expected one. Furthermore,  the data analysis will be extended in order to evaluate the null shift second order terms $\omega_{ns2}$ based on the recipes given in the present paper and to determine the associated correction in data from GINGERINO. Finally, we plan to devote efforts aimed to better understand the influence of each parameter, in particular of the change of wavelength, in the RLG operation, which for the present set of data matters, since GINGERINO is free running and the wavelength can change with even small temperature variations. 

\bibliographystyle{unsrt}

%
\pagebreak


\section*{Supplementary material}

# Appendix A: Identification of He-Ne Ring Laser dynamics

## Reference papers:

---

## Constants and parameters of He-Ne Ring Lasers

Beam − = Cw , Beam + = Ccw , Isotope 1 = Ne20, Isotope 2 = Ne22

In[1]:=
```
c =  299 792 458 (* speed of light m/s *);
h = 6.62606896 × 10^−34(* Planck constant Js *);
ε0 = 8.854187817 × 10^−12 (* Fundamental dieletric constant F/m *);
λ = 632.81 × 10^−9 (* He-Ne laser wavelength m *);
kB = 1.3806488 × 10^−23 (* Boltzmann constant J/K *);
u = 1.660538921 × 10^−27(* unified atomic mass unit kg *);
SF = L / λ / 4(* Scale Factor of the instrument;
 L perimeter length *);
FSR = c / L (* Free spectral range Hz *);
fS = X * SF (* Sagnac signal bias;
 X is the angular velocity bias of the RLG*);
```

In[10]:=
```
γa = (8.35 + p) 10^6 (* Decay rate of level a *);
γb = (9.75 + 40 p) 10^6 (* Decay rate of level b *);
Aik = 3.39 × 10^6(* Transition rate between laser levels *);
γab = (γa + γb) / 2 + δγ
  (* Radiation decay rate plus collision induced rate *);
δγ = 0;
m20 = 20 u (* atomic mass isotope 1 *);
m22 = 22 u (* atomic mass isotope 2 *);
k20 = 1 / 2 (* Relative concentration of isotope 1 *);
k22 = 1 / 2 (* Relative concentration of isotope 2 *);
```

In[19]:=
$$\Gamma D20 = \frac{\sqrt{2\,kB\,Tp\,/\,m20}}{\lambda}$$ (* Doppler halfwidth isotope 1 rad/s;

  Tp plasma temperature*);

$$\Gamma D22 = \frac{\sqrt{2\,kB\,Tp\,/\,m22}}{\lambda}$$ (* Doppler halfwidth isotope 1 rad/s *);

$$\mu ab = \sqrt{\pi\,\varepsilon0\,Aik\left(\frac{\lambda}{2\,\pi}\right)^3 \frac{h}{2\,\pi}}$$  (* Electric dipole moment of the transition *);

```
η20 = γab / ΓD20 (* homogeneus to doppler broadening rate isotope 1 *);
η22 = γab / ΓD22 (* homogeneus to doppler broadening rate isotope 2 *);
Sh = 885 × 10^6 (* isotopic shift Hz *);
```

In[25]:= **ξ120 = (0 + Sh / 2) / ΓD20**
 **(\* Detuning normalized to Doppler width for Clockwise beam, isotope 1 \*);**
 **ξ220 = (fS + Sh / 2) / ΓD20** **(\* Detuning normalized to**
 **Doppler width for CounterClockwise beam, isotope 1 \*);**
 **ξ122 = (0 - Sh / 2) / ΓD22** **(\* Detuning normalized to**
 **Doppler width for Clockwise beam, isotope 2 \*);**
 **ξ222 = (fS - Sh / 2) / ΓD22** **(\* Detuning normalized to Doppler**
 **width for CounterClockwise beam, isotope 2 \*);**
 **cI2P = $\dfrac{\pi^2\,\mu ab^2}{2\,h^2\,\gamma a\,\gamma b}\ \dfrac{1}{2\,c\,\varepsilon 0\,T\,a}$ (\* Power Calibration Constant from S.I. to Lamb Units;**
 **T Mirror transimssion coefficient, a area of the mode\*);**

### Electronic constants to express all in function of the monobeams voltage

In[30]:= **cI2V = cI2P / GAmp / aEff**
 **(\* Acquired Voltage Calibration Constant from S.I. to Lamb Units ;**
 **GAmp Transimpedance amplifier gain;**
 **aEff photodiode quantum efficency\*) ;**

### Evaluation of output power and mean intensities from the above constants

In[31]:= **Pout = Vsampled / GAmp / aEff (\* Output power 1-10 nW\*);**
 **ImeanP = cI2P Pout (\* Check Mean intensity in Lamb units \*);**
 **ImeanV = cI2V Vsampled (\* Check Mean intensity in Lamb units \*);**

### Plasma dispersion function

In[34]:= **Zi[ξ_, η_] = $\sqrt{\pi}\ e^{-\xi^2} - 2\,\eta$;**
 **(\* Imaginary part of the plasma dispersion function \*)**
 **Zr [ξ_, η_] = -2 ξ $e^{-\xi^2}$; (\* Real part of the plasma dispersion function \*)**
 **$\mathcal{L}$[ξ_, η_] = $\dfrac{1}{1 + \left(\dfrac{\xi}{\eta}\right)^2}$** **(\* Lorenzian function \*);**

### Lamb Parameters expressions for the single isotope case

In[37]:= **βs[ξ_, η_, G_] = G $\dfrac{Zi[\xi, \eta]}{Zi[0, \eta]}$ (\*Self saturation coefficient \*);**
 **αs[ξ_, η_, G_, μ_] = βs[ξ, η, G] - μ**
 **(\* Excess of gain minus losses coefficients \*);**
 **σs[ξ_, η_, G_] = G $\dfrac{FSR}{2}\ \dfrac{Zr[\xi, \eta]}{Zi[0, \eta]}$**
 **(\*Scale Factor error coefficient due to laser physics \*);**
 **θs[ξX_, ξY_, η_, G_] = βs[ξY, η, G] $\mathcal{L}$[ξX, η] (\*Cross saturation coefficient \*);**
 **τs [ξX_, ξY_, η_, G_] = G $\dfrac{FSR}{2}\left(\dfrac{\xi X}{\eta}\right)\dfrac{Zi[\xi Y, \eta]}{Zi[0, \eta]}\,\mathcal{L}$[ξX, η]**
 **(\* Null shift error coefficient \*);**

## Lamb Parameters expressions for the double isotope case
20 and 22 are the isotopes; X and Y are the counter-propagating beams

```
In[42]:=  βd[ξ20_, ξ22_, ηu20_, ηu22_, G_] = k20 βs[ξ20, ηu20, G] + k22 βs[ξ22, ηu22, G];
          αd[ξ20_, ξ22_, ηu20_, ηu22_, G_, μ_] =
              k20 αs[ξ20, ηu20, G, μ] + k22 αs[ξ22, ηu22, G, μ];
          σd[ξ20_, ξ22_, ηu20_, ηu22_, G_] = k20 σs[ξ20, ηu20, G] + k22 σs[ξ22, ηu22, G];
          θd[ξX20_, ξX22_, ξY20_, ξY22_, ηu20_, ηu22_, G_] =
              k20 θs[ξX20, ξY20, ηu20, G] + k22 θs[ξX22, ξY22, ηu22, G];
          τd[ξX20_, ξX22_, ξY20_, ξY22_, ηu20_, ηu22_, G_] =
              k20 τs[ξX20, ξY20, ηu20, G] + k22 τs[ξX22, ξY22, ηu22, G];


In[47]:=  Gain2 = Simplify[
              NSolve[ImeanP == αd[ξ220, ξ222, η20, η22, G, μ + δμ] / βd[ξ220, ξ222, η20, η22, G],
                 G][[1, 1, 2]] /. Vsampled → PH2];
          Gain1 = Simplify[NSolve[ImeanP == αd[ξ120, ξ122, η20, η22, G, μ    ] /
                 βd[ξ120, ξ122, η20, η22, G], G][[1, 1, 2]] /. Vsampled → PH1];

In[49]:=  β1 = Simplify[βd[ξ120, ξ122, η20, η22, Gain1] /. Vsampled → PH1];
          β2 = Simplify[βd[ξ220, ξ222, η20, η22, Gain2] /. Vsampled → PH2];
          DM = Solve[β1 == β2, δμ]; DMU = δμ /. DM[[1, 1]];
          β = β1;
```

## Evaluating Lamb Parameters

```
In[53]:=  σ1 = Simplify[σd[ξ120, ξ122, η20, η22, Gain1]];
          σ2 = Simplify[σd[ξ220, ξ222, η20, η22, Gain2]];
          τ12 = Simplify[τd[ξ120, ξ122, ξ220, ξ222, η20, η22, Gain1]];
          τ21 = Simplify[τd[ξ220, ξ222, ξ120, ξ122, η20, η22, Gain2]];
```

evaluation of the corrective terms
Null Shift

```
In[57]:= Δ = Sqrt[ 8 c^2 r1 r2 Cos[2 ε] /L^2 + ωm^2 ];
       δns = Simplify[σ2 - σ1 + τ21 PH2 - τ12 PH1];

       ωns1 = - δns (ωm / 2 / Δ + 1 / 2);
       ωns2 = δns^2  (2 c^2 r1 r2 Cos[2 ε]) / ((8 c^2 r1 r2 Cos[2 ε] + L^2 ωm^2) Δ);
```

K and $\omega_K$ evaluations

```
In[61]:= K = Sqrt[a1 / a2] * c * r1 * (-Cos[ε] Sin[t ωs] + Cos[t ωs] Sin[ε]) -
          Sqrt[a2 / a1] c r2 (Cos[ε] Sin[t ωs] + Cos[t ωs] Sin[ε]);

       ωK1 = K (- ωm / (2 L Δ) - 1 / (2 L));
       ωK2 = K^2 ( (2 c^2 r1 r2 Cos[2 ε] Δ) / ((8 c^2 r1 r2 Cos[2 ε] + L^2 ωm^2)^2) );
```

mixed term

```
In[64]:= ωnsK = (δns K) / (2 Sqrt[8 c^2 r1 r2 Cos[2 ε] + L^2 ωm^2]);
```

Approximated value of $\omega_{k1}$
Assuming $\langle Cos[\epsilon]\rangle = \langle Sin[\epsilon]\rangle = MK$, where MK is a proportionality constant estimated by statistics

```
In[65]:= K = MK ( Sqrt[a1/a2] IS2 L ωm (-Cos[ε] + Sin[ε]) ) / (4 Sqrt[PH1 PH2]) - ( Sqrt[a2/a1] IS1 L ωm (Cos[ε] + Sin[ε]) ) / (4 Sqrt[PH1 PH2]);

       ωK1 = MK / L (- (ωm / 2 / Δ - 1 / 2));
```

Calculations of $\alpha_{1,2}$ and $r_{1,2}$ parameters

```
In[67]:= α1 = β (PH1 + IS1^2 / 4 / PH1) + IS1 IS2 ωm Sin[2 ε] / PH2 / 4 / (c / L);
       α2 = β (PH2 + (IS2)^2 / 4 / PH2) - IS1 IS2 ωm Sin[2 ε] / PH1 / 4 / (c / L);
       r1 = IS2 ωm / 2 / (c / L) / Sqrt[PH1 PH2] / 2;
       r2 = IS1 ωm / 2 / (c / L) / Sqrt[PH1 PH2] / 2;
```

Operation Parameters of GINGERINO

```
In[71]:= p0 = 4.66 (* Gas pressure mbar *);
        p = p0 / 1.33 (* Gas pressure Torr *);
        a = π × 0.52 × 0.73 × 10⁻⁶; (*area of the mode*)
        T = 0.35 × 10⁻⁶ (* Mirror transimssion coefficient *);
        L = 4 × 3.60 (* Perimeter of the Ring m *);
        Tp = 360 (* kinetic discharge temperature *);
        aEff = 0.4 (* photodiode quantum efficency *);
        GAmp = 10⁹ (* Transimpedance amplifier gain *);
        X = 280.4 / SF; (*mean value of the anglar velocity, rad/s*)
```

---

## Calculation of the two functions used in the paper

```
In[80]:= Funω_ns1[ωm_, ε_, PH1_, PH2_, μ_, δμ_] = Simplify[ω_ns1 /. δμ → DMU];
```

```
In[81]:= Funω_K1[ωm_, ε_, PH1_, PH2_, IS1_, IS2_, μ_, MK_] =
            Assuming[ωm > 0, Simplify[ωK /. δμ → DMU]];
```

---

# In the following a few relations are explicitly reported using the parameters of GINGERINO

In this analysis $\mu$ indicates the losses of beam 2, while $\delta\mu$ is the difference between the the two beams, losses of beam 1 are $\mu + \delta\mu$

$\delta\mu$ is perfectly deducible: $\delta\mu$ = DMU

```
In[82]:= Simplify[DMU]
```

$$Out[82]= -\frac{3.23238 \times 10^{-37} \left(2.83342 \times 10^{22} + 3.0937 \times 10^{36} \, PH1 - 3.0937 \times 10^{36} \, PH2\right) \mu}{-13.7927 + 1. \, PH1}$$

$\mu$ is a proportionality function for $\omega_{ns1}$

```
In[83]:= Funω_ns1[ωm_, ε_, PH1_, PH2_, μ_, δμ_] = Simplify[ω_ns1 /. δμ → DMU]
```

$$Out[83]= \Bigg( 0.5 \left(1.20443 \times 10^{39} - 1.77025 \times 10^{43} \, PH2 + \right.$$
$$1.28346 \times 10^{42} \, PH2^2 + PH1^2 \left(-1.28347 \times 10^{42} + 9.30544 \times 10^{40} \, PH2\right) +$$
$$\left. PH1 \left(1.77025 \times 10^{43} - 8.33485 \times 10^{36} \, PH2 - 9.30534 \times 10^{40} \, PH2^2\right)\right)$$
$$\mu \left(\omega m + \sqrt{\omega m^2 \left(1 + \frac{0.5 \, IS1 \, IS2 \, Cos[2 \, \epsilon]}{PH1 \, PH2}\right)} \right) \Bigg) \Bigg/$$
$$\Bigg( (13.7927 - 1. \, PH1)^2 \left(3.0937 \times 10^{36} - 2.24299 \times 10^{35} \, PH2\right)$$
$$\sqrt{\omega m^2 \left(1 + \frac{0.5 \, IS1 \, IS2 \, Cos[2 \, \epsilon]}{PH1 \, PH2}\right)} \Bigg)$$

$\mu$ is a proportionality function for $\delta_{ns}$

```
In[84]:= Simplify[δ_ns /. δμ -> DMU]
```

$$Out[84]= \Bigg( \left(-1.20443 \times 10^{39} + PH1^2 \left(1.28347 \times 10^{42} - 9.30544 \times 10^{40} \, PH2\right) + 1.77025 \times 10^{43} \, PH2 - \right.$$
$$1.28346 \times 10^{42} \, PH2^2 + PH1 \left(-1.77025 \times 10^{43} + 8.33485 \times 10^{36} \, PH2 + 9.30534 \times 10^{40} \, PH2^2\right) \Bigg)$$
$$\mu \Bigg) \Bigg/ \left( (13.7927 - 1. \, PH1)^2 \left(3.0937 \times 10^{36} - 2.24299 \times 10^{35} \, PH2\right)\right)$$



\end{document}